# Analytical study on magnetic component of geodesic acoustic mode

**Baoyi Xie[1,2], Lei Ye[3], Yang Chen[4], Pengfei Zhao[3], Wenfeng Guo[3] and Nong Xiang[3]**

[1]Advanced Energy Research Center, Shenzhen University, Shenzhen 518060, China

[2]Key Laboratory of Optoelectronic Devices and Systems of Ministry of Education and Guangdong Province, College of physics and Optoelectronic Engineering, Shenzhen University, Shenzhen 518060, China

[3]Institute of Plasma Physics, Chinese Academy of Sciences, Hefei 230031, China

[4]Department of Physics, University of Colorado at Boulder, Boulder 80309, USA

E-mail: byxie@szu.edu.cn

## Abstract

The magnetic components of geodesic acoustic mode (GAM) are analytically investigated under the gyrokinetic framework with both the $m=1$ and $m=2$ harmonics are considered, where $m$ is the poloidal mode number. With the quasi-neutrality condition and Ampère's law, the amplitudes of various poloidal magnetic components are derived. It is shown that both $m=1$ and $m=2$ magnetic components exist and are dominated by the cosine and sine components, respectively. In addition, it is found that the amplitudes of all magnetic components increase with respect to the ratio of plasma pressure to magnetic pressure $\beta$ and safety factor $q$. Most importantly, the amplitude of $m=1$ magnetic component is significantly enhanced due to the coupling of magnetic drift frequency with the first and second harmonics of the distribution functions, thus it can be comparable to that of $m=2$ magnetic component under certain conditions.

## 1. Introduction

Zonal flows (ZFs), which are some kind of axisymmetric electrostatic potential fluctuations with finite radial wavelengths [1], are driven by the drift wave turbulence through nonlinear Reynolds stress [2], and in turn regulate turbulent transport via flow shearing, and thus have been intensively investigated in both theories and experiments [3-7]. Geodesic acoustic modes (GAMs) are the high frequency branch of ZFs and are induced by the geodesic curvature of the magnetic field line coupled to the poloidal density perturbation in toroidal configuration [8]. GAM can also regulate the drift wave turbulence and relevant transport and hence have attracted much attention in the past decades (see Refs. 9-13 and references therein). GAMs are usually considered as an electrostatic perturbation observed in many devices [14-16]. However, in recent years, multiple tokamak devices, such as T-10 [17], DIII-D [18], TCV [19], Globus-M [20], COMPASS [21] and EAST [22] also reported the magnetic signals in the frequency range of GAM and demonstrated that these magnetic perturbations have standing wave structures with poloidal mode number $m=2$. Furthermore, bispectral analysis indicates that there is a three wave coupling interaction between GAM and electromagnetic broadband turbulence [23-24]. Therefore, it is necessary to study the magnetic characteristics of GAM in detail.

Up to now, the magnetic perturbations of GAM have been analytically investigated by several authors with different models [25-33]. By using a drift kinetic model, Zhou found that only the $m=2$ magnetic component exists due to the influence of parallel return current, while the $m=1$ magnetic component disappears [25]. Later, Wang [26] investigated the magnetic components of GAM with the gyrokinetic equation and demonstrated that the $m=2$ magnetic component is one order of amplitude larger than the $m=1$ component with the condition of $\beta \sim (k\rho_{\mathrm{i}})^2$, where $\beta$ is the ratio of plasma pressure to magnetic pressure, $k$ is the radial wave number and $\rho_i$ is the ion Larmor radius. Considering the coupling to the zeroth harmonic of the distribution function ignored in [25], Bashir [27] indicated that the $m=1$ magnetic component also exists and is proportional to $\beta$. Smolyakov [28-29] found that the coupling of electromagnetic GAM and Alfvén eigenmodes can produce Alfvén sideband modes. In addition, it is shown that the plasmas toroidal rotation [30], radial equilibrium electric field [31] and shaping effects [32] also have significant influence on the magnetic components of GAM. Lately, Huang [33] studied the magnetic component of GAM with the adiabatic electron and also found that the $m=2$ magnetic component is dominant. Although the magnetic components of GAM have been studied a lot, it has not been completely resolved. Moreover, the previous theoretical studies are not complete, for example, only the $m=1$ harmonic is considered [27], or some coupling terms are ignored [25]. To study the magnetic component of GAM more accurately, not only the $m=2$ harmonic, but also some coupling terms need to be considered. This is also the main motivation of this paper.

In the present work, we investigate the magnetic components of GAM analytically with the gyrokinetic equation and consider both the $m=1$ and $m=2$ harmonics. The amplitudes of various poloidal magnetic components are derived by using the quasi-neutrality condition and Ampère's law, which shows that both $m=1$ and $m=2$ magnetic components exist and are dominated by the cosine and sine components, respectively. In addition, it is shown that the amplitudes of various magnetic components increase with $\beta$ and $q$. In particular, we consider the coupling of magnetic drift frequency with the first and second harmonics of the distribution functions. It is found that these coupling terms are very important and can greatly enhance the amplitude of $m=1$ magnetic component. Therefore, the amplitude of $m=1$ magnetic component can be comparable to that of $m=2$ magnetic component within a certain reasonable parameter range.

The paper is organized as follows. In section 2, the theoretical model is given, where both the $m=1$ and $m=2$ harmonics are considered. In section 3, the amplitudes of various poloidal magnetic components of GAM are derived. Summary and discussions are given in section 4.

## 2. Theoretical model

The perturbed distribution function can be described as $f_\alpha = -q_\alpha \phi F_{0\alpha} / T_\alpha + h_\alpha J_{0\alpha}$ ,

where $\alpha = i, e$ denotes the ions and electrons, respectively, $q_\alpha$ is the charge number, $\phi$ is perturbed electrostatic potential, $F_{0\alpha}$ is the equilibrium distribution function, $T_\alpha$ is particles temperature, $J_{0\alpha} = J_{0\alpha}(k\rho_\alpha)$ is the zero order Bessel function, $\rho_\alpha = v_\perp / \Omega_\alpha$ is the gyroradius and $\Omega_\alpha = q_\alpha B / m_\alpha$ is the gyrofrequency. The non-adiabatic response $h_\alpha$ satisfies the following gyrokinetic equation [34]

$$\left(\omega - \omega_{d\alpha} \sin\theta + i\omega_{t\alpha} \frac{\partial}{\partial\theta}\right) h_\alpha = \frac{q_\alpha F_{0\alpha} \omega J_{0\alpha}}{T_\alpha} \left(\phi - v_\parallel A\right), \tag{1}$$

where $\omega_{d\alpha} = k\left(v_\parallel^2 + v_\perp^2/2\right)/\Omega_\alpha R$ is the magnetic drift frequency, $\omega_{t\alpha} = v_{\parallel\alpha}/qR$ is the transit frequency. The equilibrium distribution function is assumed to be $F_{0\alpha} = \left(n_0/\pi^{3/2} v_{t\alpha}^3\right) \exp\left[-\left(v_\parallel^2 + v_\perp^2\right)/v_{t\alpha}^2\right]$ with $v_{t\alpha} = \sqrt{2T_\alpha / m_\alpha}$ . $A$ is the perturbation of the parallel magnetic vector potential. We consider a circular cross section as usual and work in toroidal coordinate $(r, \theta, \varphi)$, where $\theta$ and $\varphi$ are poloidal and toroidal angles, respectively. Note that $R = R_0(1 + \varepsilon \cos\theta)$, where $\varepsilon = r / R_0$ is the inverse aspect ratio. Thus the $\theta$ dependence of the transit frequency $\omega_{t\alpha}$ via the major radius $R$ may lead to the $\varepsilon$ correction. In addition, if the trapping effect is considered [35], the parallel velocity and magnetic drift velocity will also be related to $\varepsilon$. These $\varepsilon$ corrections may have a large influence on the magnetic components of GAM. However, it is very complicated to consider these $\varepsilon$ corrections in theoretical analyses. Here, for simplicity, we only consider the passing particles and lowest order terms. The equation (1) can be divided into the principal and oscillating components

$$\omega h_{\alpha 0} - \left\langle \omega_{d\alpha} \sin\theta \tilde{h}_\alpha \right\rangle = \frac{q_\alpha F_{0\alpha} J_{0\alpha}}{T_\alpha} \omega\left(\phi_0 - v_\parallel A_0\right), \tag{2}$$

$$\begin{aligned} &\omega \tilde{h}_\alpha - \omega_{d\alpha} \sin\theta h_{\alpha 0} - \left(\omega_{d\alpha} \sin\theta \tilde{h}_\alpha - \left\langle \omega_{d\alpha} \sin\theta \tilde{h}_\alpha \right\rangle\right) + i\omega_{t\alpha} \partial_\theta \tilde{h}_\alpha \\ &= \frac{q_\alpha F_{0\alpha} J_{0\alpha}}{T_\alpha} \omega\left(\tilde{\phi} - v_\parallel \tilde{A}\right), \end{aligned} \tag{3}$$

where the notation $\langle\ \rangle$ represents the average in $\theta$. In order to compare the $m = 1$

and $m=2$ magnetic components, the perturbations need to be expanded to the $m=2$ harmonic

$$\begin{aligned}X &= X_0 + X_1 e^{i\theta} + X_{-1} e^{-i\theta} + X_2 e^{2i\theta} + X_{-2} e^{-2i\theta} \\ &= X_0 + X_{1c}\cos\theta + iX_{1s}\sin\theta + X_{2c}\cos 2\theta + iX_{2s}\sin 2\theta,\end{aligned} \tag{4}$$

where $X_{1c,s} = X_1 \pm X_{-1}$, $X_{2c,s} = X_2 \pm X_{-2}$.

The equations (2) -(3) can then be simplified as

$$\omega h_{\alpha 0} - \frac{i\omega_{d\alpha}}{2} h_{\alpha 1s} = \frac{q_\alpha F_{0\alpha} J_{0\alpha}}{T_\alpha} \omega\left(\phi_0 - v_\parallel A_0\right), \tag{5}$$

$$\omega h_{\alpha 1c} - \omega_{t\alpha} h_{\alpha 1s} - \frac{i\omega_{d\alpha}}{2} h_{\alpha 2s} = \frac{q_\alpha F_{0\alpha} J_{0\alpha}}{T_\alpha} \omega\left(\phi_{1c} - v_\parallel A_{1c}\right), \tag{6}$$

$$\omega h_{\alpha 1s} - \omega_{t\alpha} h_{\alpha 1c} + i\omega_{d\alpha} h_{\alpha 0} - \frac{i\omega_{d\alpha}}{2} h_{\alpha 2c} = \frac{q_\alpha F_{0\alpha} J_{0\alpha}}{T_\alpha} \omega\left(\phi_{1s} - v_\parallel A_{1s}\right), \tag{7}$$

$$\omega h_{\alpha 2c} - 2\omega_{t\alpha} h_{\alpha 2s} + \frac{i\omega_{d\alpha}}{2} h_{\alpha 1s} = \frac{q_\alpha F_{0\alpha} J_{0\alpha}}{T_\alpha} \omega\left(\phi_{2c} - v_\parallel A_{2c}\right), \tag{8}$$

$$\omega h_{\alpha 2s} - 2\omega_{t\alpha} h_{\alpha 2c} + \frac{i\omega_{d\alpha}}{2} h_{\alpha 1c} = \frac{q_\alpha F_{0\alpha} J_{0\alpha}}{T_\alpha} \omega\left(\phi_{2s} - v_\parallel A_{2s}\right). \tag{9}$$

It is not easy to directly solve equations (5)-(9) to get accurate and complete solutions. Note that $\omega_{d\alpha}/\omega \approx k\rho_\alpha << 1$ is a reasonable assumption for GAM. Thus one can ignore the last term of the left hand of the equations (5)-(7), namely, the coupling to the first and second harmonics of the distribution functions, as done in Ref. 25, and then get the distribution functions iteratively. However, as will be discussed later, these coupling terms are very important to the magnetic components and can not be simply neglected. Here, for accuracy and completeness, we solve equations (5)-(9) directly to obtain the various distribution function components (The results are given in the appendix A due to its lengthy forms). In addition, the magnetic component of $m=0$ is far smaller than other components, so we generally do not consider it, and take $A_0 = 0$ directly.

Using the definition of the density perturbation $\delta n_\alpha = \int f_\alpha d^3 v$ and the parallel current density perturbation $\delta j_{\parallel\alpha} = q_\alpha \int f_\alpha v_\parallel d^3 v$ , the various components of the density and current density perturbations can be obtained. One can then get the various magnetic components and the general dispersion relation of GAM with the electromagnetic effects by using the quasi-neutrality condition and Ampère's law. It should be pointed out that it is too difficult to integrate the various distribution function components directly. Fortunately, we can assume that electrons are in the adiabatic regime ( $\omega qR/v_{te} < 1$ ) and ions are in the fluid regime ( $\omega qR/v_{ti} > 1$ ) , and ignore the toroidal resonance effect ( $\omega_{d\alpha}^2 < \omega^2$ ), as done in Ref. 27. Then the various density and current density perturbations in term of the generalized plasma dispersion functions can be obtained. Simplifying these equations by expanding the generalized plasma dispersion functions and keeping only the leading order terms, one can get the various components of the density and current density perturbations, as shown in the appendix B. Then with the help of the quasi-neutrality condition and Ampère's law, one can obtain the various magnetic components and the general dispersion relation of GAM with the electromagnetic effects. Here we only focus on the various magnetic components of GAM.

## 3. Derivation of magnetic components of GAM

Following the procedure outlined above, we get the cosine component of the $m = 1$ magnetic component

$$A_{1c} = -\frac{1}{S\omega qR}\left\{\left(\Gamma_{i0} - 1 + \frac{3}{4}K_2\right)\phi_{1s} + \frac{i}{2}\left(K_1 + 4K_{1\parallel}\right)\left(\phi_{2c} - \frac{1}{2}\omega qRA_{2s}\right)\right\}, \tag{10}$$

where $S = k^2\lambda_{Di}^2 / \omega^2 q^2 R^2$ . Since the frequency of GAM is $\omega^2 \approx v_{ti}^2 / R^2$ , then $S \approx k^2\lambda_{Di}^2 / q^2 v_{ti}^2 = (k\rho_i)^2 / 2\beta_i q^2$ , where $\beta_i = 2\mu_0 n_0 T_i / B^2$ , $\rho_i = v_{ti} / \Omega_i$ . When the $m = 2$ harmonic is not considered, as done in Ref. 27, then the equation (10) is reduced to $A_{1c} = -(\Gamma_{i0} - 1 + K_2/2)\phi_{1s} / S\omega qR$ , which is the same as equation (39) in Ref. 27 (A higher-order term is ignored here). Note that the term $3K_2/4$ and the terms related to $m=2$ harmonic in equation (10) are induced by the coupling of magnetic drift frequency with the first and second harmonics of the distribution functions in equations (5)-(7). Otherwise, these terms will not exist and the resultant equation (10) will become $A_{1c}=-(\Gamma_{i0}-1)\phi_{1s}/S\omega qR$. Meanwhile, if one use the drift kinetic model, i.e. the ion Larmor radius effect is ignored, then $\Gamma_{i0}\approx 1$ and the $m=1$ magnetic component is equal to zero (The $m=1$ magnetic component is dominated by the cosine component, as will be discussed later), this is the reason why only the $m=2$ magnetic component exists in Ref. 25. In addition, it can be seen from equation (10) that the $m=2$ component of the perturbed potential can enhance the $m=1$ magnetic component. These analyses indicate that the $m=2$ harmonic should be considered, and the coupling of magnetic drift frequency with the first and second harmonics of the distribution functions in equations (5)-(7) are also very important and should not be ignored.

Similarly, the other magnetic components are obtained as

$$A_{1s}=-\frac{1}{S\omega qR}\left\{\left(\Gamma_{i0}-1+\frac{1}{4}K_2\right)\phi_{1c}+\frac{i}{2}\left(K_1+4K_{1\parallel}\right)\left(\phi_{2s}-\frac{1}{2}\omega qRA_{2c}\right)\right\}, \tag{11}$$

$$A_{2c}=-\frac{1}{2S\omega qR}\left\{\left(\Gamma_{i0}-1+\frac{1}{4}K_2\right)\phi_{2s}-\frac{i}{2}\left(K_1+K_{1\parallel}\right)\left(\phi_{1c}-\omega qRA_{1s}\right)\right\}, \tag{12}$$

$$A_{2s}=-\frac{1}{2S\omega qR}\left\{\left(\Gamma_{i0}-1+\frac{1}{4}K_2\right)\phi_{2c}-\frac{1}{2}\left(K_2+K_{2\parallel}\right)\phi_0-\frac{i}{2}\left(K_1+K_{1\parallel}\right)\left(\phi_{1s}-\omega qRA_{1c}\right)\right\}. \tag{13}$$

Due to the couplings of the various components of the magnetic and electrostatic potentials, it is still difficult to get the amplitudes of various magnetic components from the equations (10)-(13). Nevertheless, we can solve it with ordering analysis. According to the definition $K_1=\varpi_{di}/\omega=kv_{ti}^2/R\Omega_i\omega$, since the GAM frequency is $\omega^2\approx v_{ti}^2/R^2$, then $K_1\approx kv_{ti}/\Omega_i=k\rho_i$, $K_2=7\varpi_{di}^2/4\omega^2\approx(k\rho_i)^2$. Meanwhile, for the general case of GAM, $K_{\parallel}=1/2\zeta_i^2\approx(k\rho_i)^2$, $\phi_m=(k\rho_i)^m\phi_0$ for $k\rho_i<<1$, thus $K_{1\parallel}=\varpi_{di}/\omega\zeta_i^2\approx(k\rho_i)^3$, $K_{2\parallel}=23\varpi_{di}^2/8\omega^2\zeta_i^2\approx(k\rho_i)^4$. Equations (10)-(13) can be greatly simplified with the help of these relations.

From the quasi-neutrality condition (see the appendix B), using the previous ordering analyses and ignoring the higher order terms, one can get the various electrostatic potential components

$$\phi_{1c} \approx \frac{i}{2}\tau K_1 \phi_{2s} + \omega q R A_{1s}, \tag{14}$$

$$\phi_{1s} \approx -i\tau K_1 \phi_0 + \frac{i}{2}\tau K_1 \phi_{2c} + \omega q R A_{1c}, \tag{15}$$

$$\phi_{2c} \approx -\frac{1}{2}\tau K_2 \phi_0 - \frac{i}{2}\tau K_1 \phi_{1s} + \frac{1}{2}\omega q R A_{2s}, \tag{16}$$

$$\phi_{2s} \approx -\frac{i}{2}\tau K_1 \phi_{1c} + \frac{1}{2}\omega q R A_{2c}. \tag{17}$$

It is found that the $m=1$ harmonic of the electrostatic potential is dominated by the sine component and $\phi_{1s} \propto (k\rho_i)\phi_0$, whereas the $m=2$ harmonic of the electrostatic potential is dominated by the cosine component and $\phi_{2c} \propto (k\rho_i)^2 \phi_0$. This is consistent with the previous results [12, 26, 33]. Using the ordering assumption and electrostatic potential relations and ignoring the higher order terms, one can get the amplitudes of the various magnetic components

$$A_{1c} \approx \frac{\beta_i q^2}{2\omega q R}\tau k\rho_i(\tau+5)\phi_0, \tag{18}$$

$$A_{1s} \approx \frac{\beta_i q^2}{\omega q R}\frac{1}{k\rho_i}\phi_{2s}, \tag{19}$$

$$A_{2c} \approx \frac{\beta_i q^2}{2\omega q R}\frac{1}{k\rho_i}\phi_{1c}, \tag{20}$$

$$A_{2s} \approx \frac{(\tau+7/4)\beta_i q^2}{2\omega q R}\phi_0. \tag{21}$$

Equations (18)-(21) clearly demonstrate that the amplitudes of various magnetic components depend on many parameters. The amplitudes of all magnetic components are proportional to $\beta$ and $q$, and increase linearly with $\beta$ and $q$. In addition, the $m=1$ and $m=2$ magnetic components both have sine and cosine components. However, from equations (14)-(17), one finds that $\phi_{1c}, \phi_{2s} \propto \mathrm{O}(k^3\rho_i^3)$, thus the $m=1$ magnetic component is dominated by the cosine component, while the $m=2$ magnetic component is dominated by the sine component, which is consistent with the previous theoretical results [25-27, 33].

From equation (4), it is shown that $A_1=(A_{1c}+A_{1s})/2$, $A_2=(A_{2c}+A_{2s})/2$. We can readily get the amplitudes of the $m=1$ and $m=2$ magnetic components

$$A_1 \approx \frac{\beta_i q^2}{4\omega qR}\tau k\rho_i(\tau+5)\phi_0, \tag{22}$$

$$A_2 \approx \frac{(\tau+7/4)\beta_i q^2}{4\omega qR}\phi_0. \tag{23}$$

Equation (22) shows that the amplitude of $m=1$ magnetic component is proportional to $k\rho_i$, which is consistent with the previous results [26-27, 33]. It is noted that the amplitude of $m=1$ magnetic component shown in equation (22) is about $(\tau+5)/2$ times larger than the results obtained in Ref. 27 where only the $m=1$ harmonic is considered. The enhancement of $m=1$ magnetic component is induced by the coupling of magnetic drift frequency with the first and second harmonics of the distribution functions in equations (5)-(7). If these couplings terms are not considered, the amplitude of $m=1$ magnetic component will much smaller than that of $m=2$ magnetic component even when $k\rho_i$ and $\tau$ are relatively large. However, after considering these couplings terms, the amplitude of $m=1$ magnetic component is significantly enhanced and thus it can be comparable to that of $m=2$ magnetic component when $k\rho_i$ and $\tau$ are relatively large, which is contrary to previous results [25-27, 33].

It should be pointed out that the equation (23) is exactly the same as the equation (38) in [33], because they are both arise from the $m=0$ and $m=1$ harmonics of the electrostatic potential. In addition, the equation (22) seems similar to the equation (34) in [33]. However, these two equations are very different. The equation (22) is derived from equation (10), which shows that the amplitude of $m=1$ magnetic component is contributed by the $m=1$ and $m=2$ harmonics of the electrostatic potential, while the electrostatic potential of $m=0$ cancels each other out, as shown in equations (B20) and (B23), thus the amplitude of $m=1$ magnetic component is proportional to the square of $\tau$. However, the equation (34) in [33] is derived from equation (17) in [33]. The amplitude of $m=1$ magnetic component is contributed by the electrostatic potential of $m=0$ and $m=1$. Therefore, the amplitude of $m=1$ magnetic component is proportional to $\tau$. When $\tau$ is relatively large, the result of equation (22) will be much larger than that of equation (34) in [33]. Moreover, it should be noted that if the electrostatic potential of $m=0$ in equation (B23) still exists, the amplitude of $m=1$ magnetic component may be much larger than that of $m=2$ magnetic component. The method used in Ref. 33 may suffer from the singularity problem [10, 27], but this problem is avoided in our work. It is not known if this is the reason why equation (22) in the present work and equation (34) in [33] are different, which requires further study.

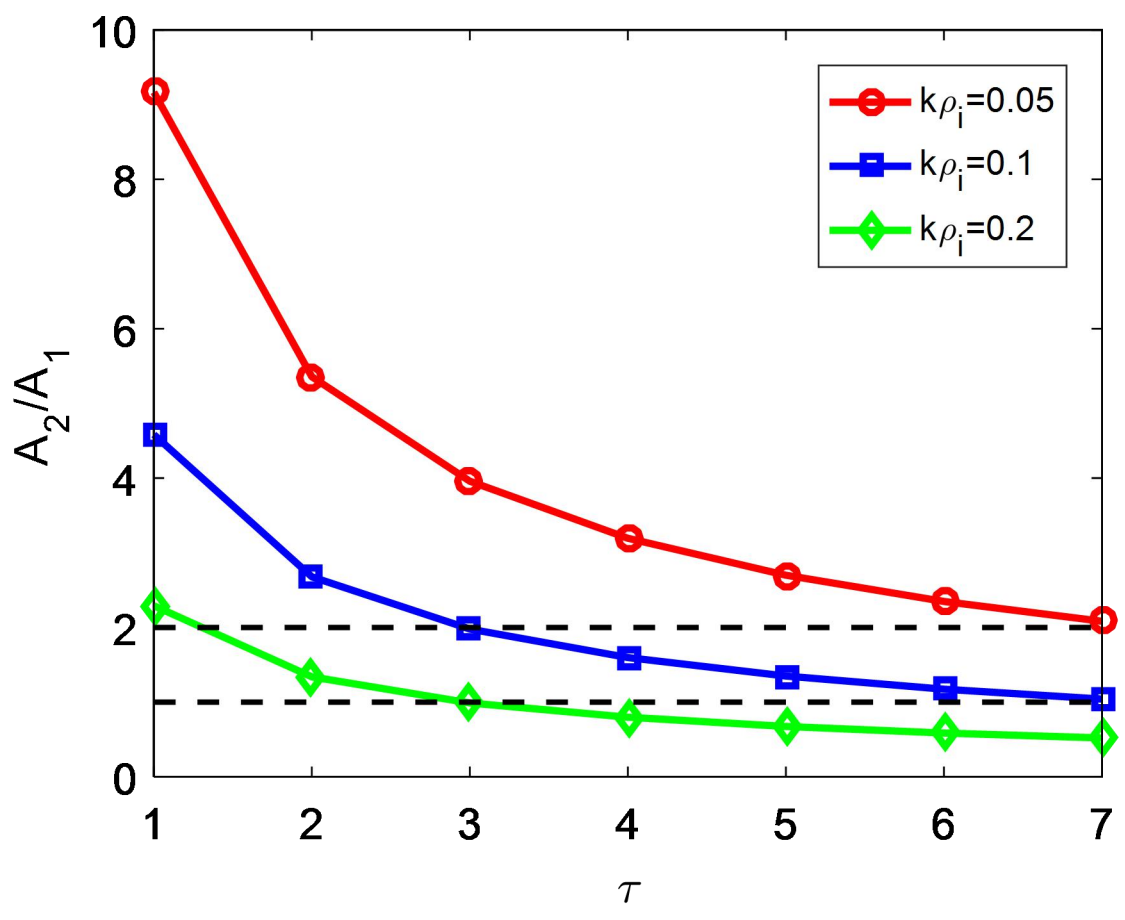

Figure 1. The ratio of $m=2$ magnetic component to the $m=1$ magnetic component versus $\tau$ with different $k\rho_i$. The two black dotted lines indicate that the ratio of $m=2$ magnetic component to the $m=1$ magnetic component is 1 and 2, respectively.

Figure 1 shows the ratio of $m=2$ magnetic component to the $m=1$ magnetic component versus $\tau$ for different $k\rho_i$. Note that $k\rho_i$ in several experimental measurements of GAM ranges from 0.04 to 0.2 [36-38]. It is found that the ratio of $m=2$ magnetic component to the $m=1$ magnetic component decreases with the increase of $\tau$ and $k\rho_i$. The amplitude of $m=1$ magnetic component is much smaller that of $m=2$ magnetic component for $k\rho_i=0.05$, even when $\tau$ is relatively large. However, when $k\rho_i=0.1$ and $\tau\geq 3$, the amplitude of $m=1$ magnetic component is comparable to that of $m=2$ magnetic component. If $k\rho_i=0.2$, the $m=1$ magnetic component will be comparable to the $m=2$ magnetic component in a wider parameter range where $\tau\geq 1$. Moreover, when $\tau\geq 3$, the $m=1$ magnetic component will be greater than $m=2$ magnetic component. These results clearly demonstrate that the magnetic components of $m=1$ and $m=2$ can be comparable within a certain reasonable parameter range. It can be seen from equations (10)-(13) that the $m=1$ and $m=2$ magnetic components are both related to $K_1$ and/or $K_2$, which are induced by the coupling of ion magnetic drift frequency and different harmonic components in equations (5)-(9). Therefore, if we do not consider these couplings terms, the amplitudes of $m=1$ and $m=2$ magnetic components will greatly decrease or even close to zero. It should be pointed out that high-order harmonic components, such as $m=3$ harmonic, may also have a great impact on the amplitudes of $m=1$ and $m=2$ magnetic components, but it is too difficult to solve in theoretical analyses, and we will not consider it here. In addition, the toroidal resonance effect is ignored in the theoretical analyses. The toroidal resonance effect has significant influence on the damping rate [12], and may also have a large effect on the magnetic components. Nevertheless, compared with the previous theoretical results [25-27, 33], our present theoretical results are more complete and accurate. The results indicate that both $m=1$ and $m=2$ magnetic components exist and can be comparable within a certain reasonable parameter range.

## 4. Summary and discussions

The various poloidal magnetic components of GAM are studied analytically with both the $m=1$ and $m=2$ harmonics are considered. By directly solving a set of gyrokinetic equations (5)-(9), we obtain complete and accurate various distribution function components (see the appendix A). The various components of the density and current density perturbations are obtained based on some approximations (see the appendix B), as done in Ref. 27. Then with the quasi-neutrality condition and Ampère's law, the amplitudes of various poloidal magnetic components are derived (see the equations (18)-(23)). The $m=1$ and $m=2$ magnetic components all exist and both have sine and cosine components. The $m=1$ magnetic component is dominated by the cosine component, while the $m=2$ magnetic component is dominated by the sine component. It is also found that the amplitudes of various magnetic components increase with $\beta$ and $q$. In addition, the amplitude of $m=1$ magnetic component is proportional to $k\rho_i$ and the square of $\tau$, thus it is much smaller than that of $m=2$ magnetic component when $k\rho_i$ and $\tau$ are relatively small. These are consistent with the previous theoretical results [25-27, 33].

It is noted that the amplitude of $m=1$ magnetic component is greatly enhanced due to the coupling of magnetic drift frequency with the first and second harmonics of the distribution functions, and it is about $(\tau+5)/2$ times larger than the results obtained in Ref. 27 where only the $m=1$ harmonic is considered. Therefore, the amplitude of $m=1$ magnetic component can be comparable to that of $m=2$ magnetic component when $k\rho_i$ and $\tau$ are relatively large, which is contrary to previous results [25-27, 33]. However, the present results seem more consistent with some experimental results that magnetic signals in the frequency range of GAM are more complicated than a single $m$ due to the anti-ballooning structure observed on multiple tokamak devices [18-19, 21]. Recently, a review on GAMs also reports the detection of $m=2$ and/or $m>2$ components [39]. These experimental results indicate that there may be multiple poloidal components in the magnetic signals of GAM. It should be pointed out that many approximations (such as the neglect of higher order harmonic coupling and toroidal resonance effect and so on) have been made in the theoretical derivations. Nevertheless, the expressions of various magnetic components derived in the present work can provide certain guiding significance for the experiments. To fully understand the magnetic component of GAM, more detailed experiments and more accurate simulations are needed, which are left for future study. In addition, some experimental results show that the magnetic component of GAM may play an important role in regulating electromagnetic turbulence [23-24], but this is beyond the scope of this paper and will be studied in the future.

## Acknowledgment

This work is supported by the National MCF Energy R&D Program of China under No. 2019YFE03060000 and the National Key R&D Program of China under Grant No. 2017YFE0300400 and the National Natural Science Foundation of China under Grant Nos. 12105188 and 11775268. This work is also supported by Shenzhen Clean Energy Research Institute.

## Appendix A.

The various distribution function components are obtained as

$$\begin{aligned} h_{\alpha 1c} = \frac{q_\alpha F_{0\alpha} J_{0\alpha}}{T_\alpha} \frac{2}{W} \Big\{ & -4i\omega_{t\alpha}\omega_{d\alpha}\left(2\omega^2 - 8\omega_{t\alpha}^2 + \omega_{d\alpha}^2\right)\phi_0 + \left(8\omega^4 - 32\omega^2\omega_{t\alpha}^2 \right. \\ & \left. -6\omega^2\omega_{d\alpha}^2 + 16\omega_{t\alpha}^2\omega_{d\alpha}^2\right)\left(\phi_{1c} - v_\| A_{1c}\right) + 4\omega\omega_{t\alpha}\left(2\omega^2 - 8\omega_{t\alpha}^2 + \omega_{d\alpha}^2\right)\left(\phi_{1s} - v_\| A_{1s}\right) \\ & + 4i\omega_{t\alpha}\omega_{d\alpha}\left(3\omega^2 - \omega_{d\alpha}^2\right)\left(\phi_{2c} - v_\| A_{2c}\right) + i\omega\omega_{d\alpha}\left(4\omega^2 - 3\omega_{d\alpha}^2 + 8\omega_{t\alpha}^2\right)\left(\phi_{2s} - v_\| A_{2s}\right)\Big\}, \end{aligned} \tag{A1}$$

$$\begin{aligned} h_{\alpha 1s} = \frac{q_\alpha F_{0\alpha} J_{0\alpha}}{T_\alpha} \frac{2\omega}{W} \Big\{ & -2i\omega_{d\alpha}\left(4\omega^2 - 16\omega_{t\alpha}^2 - \omega_{d\alpha}^2\right)\phi_0 + 4\omega_{t\alpha}\left(2\omega^2 - 8\omega_{t\alpha}^2 + \omega_{d\alpha}^2\right) \\ & \times\left(\phi_{1c} - v_\| A_{1c}\right) + 2\omega\left(4\omega^2 - 16\omega_{t\alpha}^2 - \omega_{d\alpha}^2\right)\left(\phi_{1s} - v_\| A_{1s}\right) + i\omega_{d\alpha}\left(4\omega^2 - \omega_{d\alpha}^2 + 8\omega_{t\alpha}^2\right) \\ & \times\left(\phi_{2c} - v_\| A_{2c}\right) + 12i\omega\omega_{t\alpha}\omega_{d\alpha}\left(\phi_{2s} - v_\| A_{2s}\right)\Big\}, \end{aligned} \tag{A2}$$

$$\begin{aligned} h_{\alpha 2c} = \frac{q_\alpha F_{0\alpha} J_{0\alpha}}{T_\alpha} \frac{2}{W} \Big\{ & -\omega_{d\alpha}^2\left(4\omega^2 - \omega_{d\alpha}^2 + 8\omega_{t\alpha}^2\right)\phi_0 - 4i\omega_{t\alpha}\omega_{d\alpha}\left(3\omega^2 - \omega_{d\alpha}^2\right) \\ & \times\left(\phi_{1c} - v_\| A_{1c}\right) - i\omega\omega_{d\alpha}\left(4\omega^2 - \omega_{d\alpha}^2 + 8\omega_{t\alpha}^2\right)\left(\phi_{1s} - v_\| A_{1s}\right) + \left(8\omega^4 - 8\omega^2\omega_{t\alpha}^2 \right. \\ & \left. -6\omega^2\omega_{d\alpha}^2 + \omega_{d\alpha}^4\right)\left(\phi_{2c} - v_\| A_{2c}\right) + 2\omega\omega_{t\alpha}\left(8\omega^2 - 8\omega_{t\alpha}^2 - 3\omega_{d\alpha}^2\right)\left(\phi_{2s} - v_\| A_{2s}\right)\Big\}, \end{aligned} \tag{A3}$$

$$\begin{aligned} h_{\alpha 2s} = \frac{q_\alpha F_{0\alpha} J_{0\alpha}}{T_\alpha} \frac{2\omega}{W} \Big\{ & -12\omega_{t\alpha}\omega_{d\alpha}^2\phi_0 - i\omega_{d\alpha}\left(4\omega^2 - 3\omega_{d\alpha}^2 + 8\omega_{t\alpha}^2\right)\left(\phi_{1c} - v_\| A_{1c}\right) \\ & -12i\omega\omega_{t\alpha}\omega_{d\alpha}\left(\phi_{1s} - v_\| A_{1s}\right) + 2\omega_{t\alpha}\left(8\omega^2 - 8\omega_{t\alpha}^2 - 3\omega_{d\alpha}^2\right)\left(\phi_{2c} - v_\| A_{2c}\right) \\ & + 2\omega\left(4\omega^2 - 4\omega_{t\alpha}^2 - 3\omega_{d\alpha}^2\right)\left(\phi_{2s} - v_\| A_{2s}\right)\Big\}, \end{aligned} \tag{A4}$$

where $W = 16\omega^4 - 80\omega_{t\alpha}^2\omega^2 + 64\omega_{t\alpha}^4 - 16\omega_{d\alpha}^2\omega^2 + 16\omega_{t\alpha}^2\omega_{d\alpha}^2 + 3\omega_{d\alpha}^4$ . Note that the distribution function of $m=0$ is not required here, so it is not given.

## Appendix B.

It is very difficult to integrate the various distribution functions directly, so we need to do some approximations, as done in Ref. 27. For the electrons, one can assume $\omega_{de}^2<\omega^2$ and $J_0^2\left(kv_\perp/\Omega_e\right)\approx 1$, and ignore the toroidal coupling effect, the expression of the denominator of the distribution function can be simplified to

$$W=16\left(\omega^2-\omega_{te}^2\right)\left(\omega^2-4\omega_{te}^2\right) \quad \text{(B1)}$$

While for the ions, we can assume $\omega_{di}^2,\omega_{ti}^2<\omega^2$, and ignore the higher order terms, the expression can be written as

$$\frac{1}{W}\approx\frac{1}{16\omega^4}\left(1+5\frac{\omega_{ti}^2}{\omega^2}+\frac{\omega_{di}^2}{\omega^2}+9\frac{\omega_{ti}^2\omega_{di}^2}{\omega^4}\right) \quad \text{(B2)}$$

Using these approximations, one can get the various density and current density perturbations in term of the generalized plasma dispersion functions. Simplifying these equations by expanding the various generalized plasma dispersion functions and keeping only the leading order terms, the simplified density and current density perturbations can be finally obtained.

The simplified electron density and current density perturbations are as follows

$$\delta n_{e1c}=-\frac{q_e n_{0e}}{T_e}\left(\phi_{1c}-\omega qRA_{1s}\right), \quad \text{(B3)}$$

$$\delta n_{e1s}=-\frac{q_e n_{0e}}{T_e}\left(\phi_{1s}-\omega qRA_{1c}\right), \quad \text{(B4)}$$

$$\delta n_{e2c}=-\frac{q_e n_{0e}}{T_e}\left(\phi_{2c}-\frac{1}{2}\omega qRA_{2s}\right), \quad \text{(B5)}$$

$$\delta n_{e2s}=-\frac{q_e n_{0e}}{T_e}\left(\phi_{2s}-\frac{1}{2}\omega qRA_{2c}\right), \quad \text{(B6)}$$

$$\delta J_{e1c}=-\frac{q_e^2 n_{0e}}{T_e}\omega qR\left(-i\frac{\varpi_{de}}{\omega}\phi_0+\phi_{1s}-\omega qRA_{1c}+\frac{i}{4}\frac{\varpi_{de}}{\omega}\omega qRA_{2s}\right), \quad \text{(B7)}$$

$$\delta J_{e1s} = -\frac{q_e^2 n_{0e}}{T_e}\omega qR\left(\phi_{1c} - \omega qRA_{1s} + \frac{i}{4}\frac{\overline{\omega}_{de}}{\omega}\omega qRA_{2c}\right), \tag{B8}$$

$$\delta J_{e2c} = -\frac{q_e^2 n_{0e}}{T_e}\frac{\omega qR}{2}\left(\phi_{2s} - \frac{1}{2}\omega qRA_{2c} - \frac{i}{2}\frac{\overline{\omega}_{de}}{\omega}\omega qRA_{1s}\right), \tag{B9}$$

$$\delta J_{e2s} = -\frac{q_e^2 n_{0e}}{T_e}\frac{\omega qR}{2}\left(\phi_{2c} - \frac{1}{2}\omega qRA_{2s} - \frac{i}{2}\frac{\overline{\omega}_{de}}{\omega}\omega qRA_{1c}\right), \tag{B10}$$

where $\overline{\omega}_{de} = kv_{te}^2 / R\Omega_e$ .

The simplified ion density and current density perturbations are obtained as

$$\begin{aligned}\delta n_{i1c} = \frac{q_i n_{0i}}{T_i}\Bigg\{&\left(\Gamma_{i0} - 1 + \frac{1}{4}K_2\right)\phi_{1c} + K_{\|}\left(\phi_{1c} - \omega qRA_{1s}\right) - \frac{3i}{2}K_{1\|}\omega qRA_{2c} \\ &+ i\left(\frac{1}{2}K_1 + \frac{7}{2}K_{1\|}\right)\phi_{2s}\Bigg\},\end{aligned} \tag{B11}$$

$$\begin{aligned}\delta n_{i1s} = \frac{q_i n_{0i}}{T_i}\Bigg\{&\left(\Gamma_{i0} - 1 + \frac{3}{4}K_2\right)\phi_{1s} - i\left(K_1 + K_{1\|}\right)\phi_0 + K_{\|}\left(\phi_{1s} - \omega qRA_{1c}\right) \\ &- \frac{3i}{2}K_{1\|}\omega qRA_{2s} + i\left(\frac{1}{2}K_1 + \frac{7}{2}K_{1\|}\right)\phi_{2c}\Bigg\},\end{aligned} \tag{B12}$$

$$\begin{aligned}\delta n_{i2c} = \frac{q_i n_{0i}}{T_i}\Bigg\{&\left(\Gamma_{i0} - 1 + \frac{1}{4}K_2\right)\phi_{2c} - \left(\frac{1}{2}K_2 + \frac{7}{2}K_{2\|}\right)\phi_0 + 4K_{\|}\left(\phi_{2c} - \frac{1}{2}\omega qRA_{2s}\right) \\ &+ \frac{3i}{2}K_{1\|}\omega qRA_{1c} - i\left(\frac{1}{2}K_1 + \frac{7}{2}K_{1\|}\right)\phi_{1s}\Bigg\},\end{aligned} \tag{B13}$$

$$\begin{aligned}\delta n_{i2s} = \frac{q_i n_{0i}}{T_i}\Bigg\{&\left(\Gamma_{i0} - 1 + \frac{1}{4}K_2\right)\phi_{2s} + 4K_{\|}\left(\phi_{2s} - \frac{1}{2}\omega qRA_{2c}\right) + \frac{3i}{2}K_{1\|}\omega qRA_{1s} \\ &- i\left(\frac{1}{2}K_1 + \frac{7}{2}K_{1\|}\right)\phi_{1c}\Bigg\},\end{aligned} \tag{B14}$$

$$\delta J_{i1c} = \frac{q_i^2 n_{0i}}{T_i}\omega qR\left\{-iK_{1\|}\phi_0 + K_{\|}\left(\phi_{1s} - \omega qRA_{1c}\right) + \frac{3i}{2}K_{1\|}\phi_{2c} - \frac{i}{2}K_{1\|}\omega qRA_{2s}\right\}, \tag{B15}$$

$$\delta J_{i1s} = \frac{q_i^2 n_{0i}}{T_i}\omega qR\left\{K_{\|}\left(\phi_{1c} - \omega qRA_{1s}\right) + \frac{3i}{2}K_{1\|}\phi_{2s} - \frac{i}{2}K_{1\|}\omega qRA_{2c}\right\}, \tag{B16}$$

$$\delta J_{i2c} = \frac{q_i^2 n_{0i}}{T_i}\omega qR\left\{2K_{\|}\left(\phi_{2s} - \frac{1}{2}\omega qRA_{2c}\right) - \frac{3i}{2}K_{1\|}\phi_{1c} + \frac{i}{2}K_{1\|}\omega qRA_{1s}\right\}, \tag{B17}$$

$$\delta J_{i2s} = \frac{q_i^2 n_{0i}}{T_i}\omega qR\left\{-\frac{3}{2}K_{2\|}\phi_0 + 2K_{\|}\left(\phi_{2c} - \frac{1}{2}\omega qRA_{2s}\right) - \frac{3i}{2}K_{1\|}\phi_{1s} + \frac{i}{2}K_{1\|}\omega qRA_{1c}\right\}, \tag{B18}$$

where $\Gamma_{i0}=1-(k\rho_i)^2/2$ , $K_1=\varpi_{di}/\omega$ , $K_2=7\varpi_{di}^2/4\omega^2$ , $K_{\parallel}=1/2\zeta_i^2$ , $K_{1\parallel}=\varpi_{di}/\omega\zeta_i^2$ , $K_{2\parallel}=23\varpi_{di}^2/8\omega^2\zeta_i^2$ , $\varpi_{di}=kv_{ti}^2/R\Omega_i$ , $\zeta_i=qR\omega/v_{ti}$ .

Using the equations (B3)-(B6) and (B11)-(B14), the quasi-neutrality condition can be written as follows

$$\phi_{1c}-\omega qRA_{1s}=\frac{1}{\tau^{-1}-K_{\parallel}}\left\{\left(\Gamma_{i0}-1+\frac{1}{4}K_2\right)\phi_{1c}-\frac{3i}{2}K_{1\parallel}\omega qRA_{2c}+i\left(\frac{1}{2}K_1+\frac{7}{2}K_{1\parallel}\right)\phi_{2s}\right\}, \quad \text{(B19)}$$

$$\phi_{1s}-\omega qRA_{1c}=\frac{1}{\tau^{-1}-K_{\parallel}}\left\{\left(\Gamma_{i0}-1+\frac{3}{4}K_2\right)\phi_{1s}-i\left(K_1+K_{1\parallel}\right)\phi_0-\frac{3i}{2}K_{1\parallel}\omega qRA_{2s}+i\left(\frac{1}{2}K_1+\frac{7}{2}K_{1\parallel}\right)\phi_{2c}\right\}, \quad \text{(B20)}$$

$$\phi_{2c}-\frac{1}{2}\omega qRA_{2s}=\frac{1}{\tau^{-1}-4K_{\parallel}}\left\{\left(\Gamma_{i0}-1+\frac{1}{4}K_2\right)\phi_{2c}-\left(\frac{1}{2}K_2+\frac{7}{2}K_{2\parallel}\right)\phi_0+\frac{3i}{2}K_{1\parallel}\omega qRA_{1c}-i\left(\frac{1}{2}K_1+\frac{7}{2}K_{1\parallel}\right)\phi_{1s}\right\}, \quad \text{(B21)}$$

$$\phi_{2s}-\frac{1}{2}\omega qRA_{2c}=\frac{1}{\tau^{-1}-4K_{\parallel}}\left\{\left(\Gamma_{i0}-1+\frac{1}{4}K_2\right)\phi_{2s}+\frac{3i}{2}K_{1\parallel}\omega qRA_{1s}-i\left(\frac{1}{2}K_1+\frac{7}{2}K_{1\parallel}\right)\phi_{1c}\right\}, \quad \text{(B22)}$$

where $\tau=T_e/T_i$ .

Combining equations (B7)-(B10) and (B15)-(B18), the total current density perturbations can be obtained as

$$\delta J_{1c}=-\frac{q_i^2n_{0i}}{T_i}\omega qR\left\{\left(\tau^{-1}-K_{\parallel}\right)\left(\phi_{1s}-\omega qRA_{1c}\right)+i\left(K_1+K_{1\parallel}\right)\phi_0-\frac{3i}{2}K_{1\parallel}\phi_{2c}-\frac{i}{4}\left(K_1-2K_{1\parallel}\right)\omega qRA_{2s}\right\}, \quad \text{(B23)}$$

$$\delta J_{1s}=-\frac{q_i^2n_{0i}}{T_i}\omega qR\left\{\left(\tau^{-1}-K_{\parallel}\right)\left(\phi_{1c}-\omega qRA_{1s}\right)-\frac{3i}{2}K_{1\parallel}\phi_{2s}-\frac{i}{4}\left(K_1-2K_{1\parallel}\right)\omega qRA_{2c}\right\}, \quad \text{(B24)}$$

$$\delta J_{2c}=-\frac{q_i^2n_{0i}}{T_i}\omega qR\left\{\frac{1}{2}\left(\tau^{-1}-4K_{\parallel}\right)\left(\phi_{2s}-\frac{1}{2}\omega qRA_{2c}\right)+\frac{3i}{2}K_{1\parallel}\phi_{1c}+\frac{i}{4}\left(K_1-2K_{1\parallel}\right)\omega qRA_{1s}\right\}, \quad \text{(B25)}$$

$$\delta J_{2s} = -\frac{q_i^2 n_{0i}}{T_i}\omega qR\left\{\frac{1}{2}\left(\tau^{-1} - 4K_{\|}\right)\left(\phi_{2c} - \frac{1}{2}\omega qRA_{2s}\right) + \frac{3}{2}K_{2\|}\phi_0 + \frac{3i}{2}K_{1\|}\phi_{1s} + \frac{i}{4}\left(K_1 - 2K_{1\|}\right)\omega qRA_{1c}\right\}, \tag{B26}$$

then with the Ampère's law and the quasi-neutrality condition (B19)-(B22), one can get the equations (10)-(13).